ENGINEERING

# In-air microfluidics enables rapid fabrication of emulsions, suspensions, and 3D modular (bio)materials

Claas Willem Visser,[1]*[†][‡] Tom Kamperman,[2]* Lisanne P. Karbaat,[2] Detlef Lohse,[1] Marcel Karperien[2][‡]

Microfluidic chips provide unparalleled control over droplets and jets, which have advanced all natural sciences. However, microfluidic applications could be vastly expanded by increasing the per-channel throughput and directly exploiting the output of chips for rapid additive manufacturing. We unlock these features with in-air microfluidics, a new chip-free platform to manipulate microscale liquid streams in the air. By controlling the composition and in-air impact of liquid microjets by surface tension–driven encapsulation, we fabricate monodisperse emulsions, particles, and fibers with diameters of 20 to 300 μm at rates that are 10 to 100 times higher than chip-based droplet microfluidics. Furthermore, in-air microfluidics uniquely enables module-based production of three-dimensional (3D) multiscale (bio)materials in one step because droplets are partially solidified in-flight and can immediately be printed onto a substrate. In-air microfluidics is cytocompatible, as demonstrated by additive manufacturing of 3D modular constructs with tailored microenvironments for multiple cell types. Its in-line control, high throughput and resolution, and cytocompatibility make in-air microfluidics a versatile platform technology for science, industry, and health care.

## INTRODUCTION

Microfluidics has become a cornerstone platform technology for testing and production of microscale droplets, particles, and fibers as well as encapsulation of food, drugs, and cells (1–4). This versatility originates from the use of microfluidic chips, which combine fully predictable flow behavior with in-line liquid manipulation and monitoring (5, 6). As a typical example, a coaxial chip geometry is shown in Fig. 1A. Here, liquid is ejected from the inner channel and pulled off by an outer coflowing liquid, resulting in a monodisperse train of droplets. Alternative designs and more advanced chips have been used for the production of microdroplets, microparticles, and microfibers with a wide variety of sizes, shapes, and compositions (7–10). However, despite their success regarding laboratory-scale analysis and production, conventional microfluidic chips have intrinsic limitations that hamper translation of successful concepts into clinical, pharmaceutical, or industrial products (11–13). First, microfluidic droplet generators are typically operated at a per-nozzle throughput of 1 to 10 μl/min because of a transition from monodisperse dripping to polydisperse jetting for higher flow rates (14). Second, the design, fabrication, and operation of microfluidic devices require advanced skills and specialized equipment, which are not always compatible with existing production processes or environments outside the laboratory (6, 15). Third, microfluidic chips can only be operated with at least one nonsolidifying flow, which is required to separate droplets, particles, or fibers from each other and the channel walls (3). Removing this coflow (for example, oil) is nontrivial, which limits clinical translation and implies that the output of chips is limited to particles in solution (that is, emulsions and suspensions), whereas stacking of particles into solid three-dimensional (3D) constructs could enable printing of modular materials in one step. An off-chip approach would eliminate all these wall-induced limitations and therefore unlock new applications of microfluidics.

Here, we present in-air microfluidics (IAMF), a new chip-free platform technology that enables in-flight (that is, on-the-fly) formation of droplets, fibers, and particles and their one-step deposition into 3D constructs with a modular internal architecture. In concept, microfluidic channels are replaced by micrometer-sized liquid jets that are combined in the air, as shown in Fig. 1B. This approach retains the on-the-fly processing capacity of chip-based microfluidics but enables orders of magnitude higher throughput. In this work, we (i) discuss the physical principles that underlie IAMF; (ii) leverage surface tension–driven encapsulation to achieve in-air liquid-liquid encapsulation; (iii) exploit these mechanisms to realize a library of droplets, particles, and fibers with distinct shapes, compositions, and sizes; and (iv) demonstrate that these in-air formed particles can be used as cytocompatible "building blocks" for the one-step printing of larger 3D (bio)materials with various modular architectures. Finally, we discuss IAMF in relation to other technologies, as well as potential directions for future work.

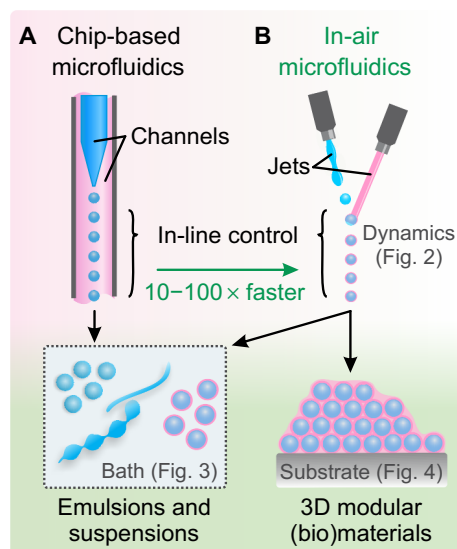

**Fig. 1. Concept of IAMF and guide to the article.** (**A**) Chip-based microfluidics enables in-line control over droplets and particles, making it a versatile platform technology. A chip design where droplets (blue) are transported by a coflow (pink) is shown. (**B**) IAMF maintains the in-line control of chip-based microfluidics but relies on jet ejection and coalescence into air. Therefore, a wide range of droplets and particles can be produced at flow rates typically two orders of magnitude higher than with chip-based microfluidics. When combining reactive, solidifying microjets, IAMF also enables on-the-fly production and direct deposition of microparticles into 3D multiscale modular (bio)materials.

[1]Physics of Fluids Group, Faculty of Science and Technology, University of Twente, P.O. Box 217, 7500 AE Enschede, Netherlands. [2]Department of Developmental BioEngineering, MIRA Institute for Biomedical Engineering and Technical Medicine, University of Twente, P.O. Box 217, 7500 AE Enschede, Netherlands.
*These authors contributed equally to this work.
†Present address: Wyss Institute for Biologically Inspired Engineering and John A. Paulson School of Engineering and Applied Sciences, Harvard University, Cambridge, MA 02138, USA.
‡Corresponding author. Email: c.visser@utwente.nl (C.W.V.); h.b.j.karperien@utwente.nl (M.K.)

## RESULTS
### Physical principles of IAMF

In IAMF, liquid microjets are manipulated and combined in the air, as shown in Fig. 2A for the "drop-jet" mode (the setup is depicted in fig. S1). Droplets are generated by breakup of the liquid jet ejected from nozzle 1. These droplets are monodisperse, as achieved by mounting the nozzle onto a vibrating piezoelectric element (section S1 and fig. S2). The droplet train impacts onto an intact liquid jet that is ejected from nozzle 2, resulting in a compound monodisperse droplet train flowing downward. While flying in the air, the compounds react chemically or physically to form encapsulated droplets, particles, or, if the setup is operated in "jet-jet" mode (Fig. 2B), fibers.

The key physical mechanisms of IAMF are in-air impact, encapsulation, and solidification. First, as sketched in Fig. 2C, a droplet impacts onto a jet. Impact must result in coalescence, whereas droplet bouncing, stretching, or splashing must be prevented (16). Furthermore, maintaining the production of spherical particles is promoted if the droplet remains spherical during impact. Both these conditions are met if capillary forces dominate inertia, that is, for impact Weber numbers $We_{impact} = \rho_1 D_D V_{impact}^2/\sigma_1 \lesssim 3$, where $\rho_1$, $\sigma_1$, and $D_D$ are the droplet density, surface tension, and diameter, respectively (16, 17). The impact velocity $V_{impact} = V_1 \sin\theta$ depends on the impact angle $\theta$ and the ejection velocity $V_1$ of jet 1. Because a significant ejection velocity is required for jet formation (see fig. S2, A to C), a small impact angle $\theta = 25 \pm 5°$ was chosen to ensure a low impact Weber number. Figure 2D, where the droplet train is selectively labeled with a fluorescent dye, shows that the deformation of the droplets is limited during in-air impact. For $We_{impact} \lesssim 1$, the coalescence of the droplets occurs at an inertial-capillary time scale $\tau_{cap} = (\rho_1 D_D^3/\sigma_1)^{1/2}$ (18).

The crucial trick that prevents the droplets ejected from nozzle 1 from merging during flight is to ensure their encapsulation by the intact jet. To drive this in-air encapsulation, the surface tension of the encapsulating (jet) liquid was reduced with respect to that of the droplet by adding ethanol. As discussed below, a surface tension difference $\Delta\sigma > 10$ mN/m (corresponding to adding 1% ethanol) was sufficient to achieve encapsulation. As a result, a Marangoni flow (that is, driven by surface tension gradients) pulls a thin film of the low surface tension liquid around the high surface tension liquid, as depicted in Fig. 2C. This mechanism allows encapsulation by both miscible (19) and immiscible (20, 21) liquids while limiting droplet deformations. The distance covered by the outer droplet is described by $L(t) = [\Delta\sigma^2 t^3/(\rho_1 \mu_1)]^{1/4}$ (22). Assuming encapsulation when $L(t) = D_D$, this relationship provides an encapsulation time scale of $\tau_e \sim [\rho_1 \mu_1 D_D^4/\Delta\sigma^2]^{1/3}$. For our experimental conditions, $\tau_e$ is comparable to the impact time scale $\tau_{cap}$. Therefore, both impact and encapsulation are completed in the air before collection or deposition, which happens typically ~100 ms after in-air impact. The robustness of in-air droplet formation for different liquid pairs is shown in fig. S2 (D to I).

Finally, the droplets could be solidified by combining two reactive liquids. To demonstrate this concept, we produced alginate-containing droplets that were solidified via fast (that is, in-air) ionotropic cross-linking by divalent $Ca^{2+}$ ions, which were added to the jet. By introducing a surface tension gradient $\Delta\sigma$, the particle shape could be tuned from irregular ($\Delta\sigma = 0$ mN/m; Fig. 2F) to spherical ($\Delta\sigma = 20$ mN/m; Fig. 2G). This indicated the consecutive occurrence of surface tension–driven encapsulation and solidification of the droplets. The particles in Fig. 2 (F and G) all had the same volume, but the nonspherical particles appeared larger because the volume was contained by a bag-like shape. Figure 2H shows the particle shape as a function of the surface tension gradient and the nozzle size. The regime transition from irregular to spherical particles was observed for $\Delta\sigma \approx 5$ mN/m, corresponding to adding a minimum amount of 0.3% ethanol.

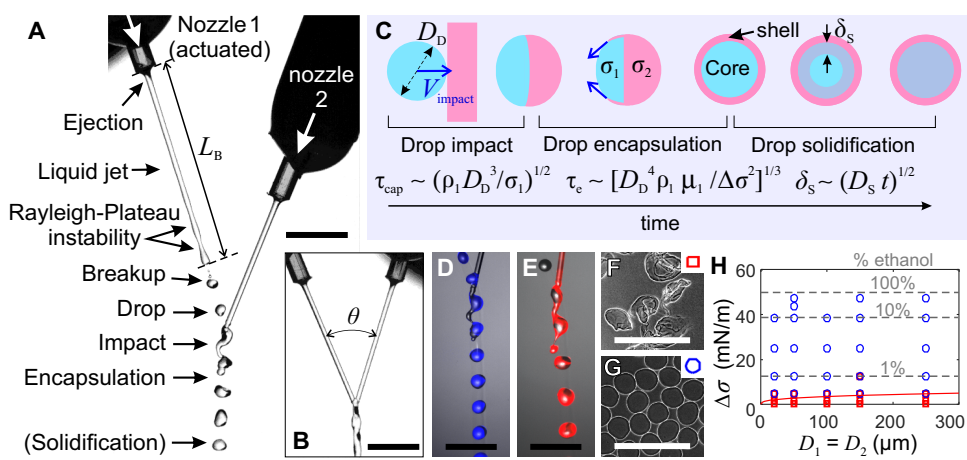

**Fig. 2. Physical principles of IAMF.** (**A**) High-speed photograph of IAMF operated in "drop-jet" mode. Here, a droplet train is ejected from actuated nozzle 1 and collides with a jet that is ejected from nozzle 2. (**B**) IAMF operated in "jet-jet" mode, as used for spinning fibers. (**C**) Schematic representation of in-air impact, encapsulation, and solidification mechanisms. Different surface tensions ($\sigma_1 > \sigma_2$) result in Marangoni-driven encapsulation of the droplet. (**D** and **E**) High-speed photographs of the "drop-jet" mode, in which (D) the droplets and (E) the jet are selectively labeled with a fluorescent dye. The droplets maintain a spherical shape during impact and encapsulation, whereas the jet spreads around the droplet within a few diameters of travel. (**F** and **G**) Alginate microparticles produced (F) without and (G) with Marangoni-driven encapsulation. (**H**) Phase diagram of particle shape as a function of surface tension gradient $\Delta\sigma$ and the nozzle diameter $D_1 = D_2$. Symbols indicate spherical (o) or irregular (□) particles. The solid line refers to a basic model for the transition between these shape regimes. Scale bars, 1 mm (black) and 0.4 mm (white).

It is surprising that the particle shape can be controlled by combining surface tension–driven encapsulation and solidification because even a thin solid front could potentially inhibit the Marangoni flow. To provide a first rationalization of mechanism, we hypothesize that encapsulation is achieved if the surface tension gradient exceeds the strength of the solidifying film. The thickness of this film is estimated as $\delta_s = (D_S\tau_e)^{1/2}$, where $D_S \approx 10^{-9}$ m$^2$ s$^{-1}$ is the diffusion constant of CaCl$_2$ into the gel (23). The strength of the film is estimated as $\sigma_f\delta_s$, where $\sigma_f \approx 10^4$ Pa is the fracture stress of a 0.5% alginate gel (24). By equating $\sigma_f\delta_s = \Delta\sigma$ and solving for $\Delta\sigma$, one obtains the solid line in Fig. 2H. For the measured parameter regime, the expected film strength lies between 2 and 5 mN/m, which is close to the experimental threshold $\Delta\sigma \approx 5$ mN/m. However, the predicted dependence on the diameter is not observed, possibly because the time-dependent viscosity gradients are ignored in our simplified model.

### Engineering droplets, particles, and fibers

Figure 3 shows droplets, particles, and fibers as produced by tuning the control parameters of IAMF. First, different material compositions were examined while operating the setup in drop-jet mode (Fig. 3A and table S1). Coalescing water droplets onto a surfactant-containing fluorocarbon oil jet (with $\Delta\sigma = 50 \pm 5$ mN/m) readily enabled the production of monodisperse water-in-oil (w/o) emulsions, as shown in Fig. 3B. Moreover, collecting these w/o droplets in surfactant-containing water resulted in w/o/w double emulsions (Fig. 3C). IAMF also enabled oil-free production of monodisperse solid particles such as alginate microspheres, as shown in Fig. 3D. Alternatively, liquid-filled microcapsules were produced by coalescing CaCl$_2$ droplets onto an alginate jet with reduced surface tension, as shown in Fig. 3E and fig. S3A. Solid-filled capsules were made by adding an in situ cross-linkable dextran-tyramine–based hydrogel precursor to the droplets and its cross-linker to the intact jet, as shown in Fig. 3F. IAMF is also compatible with slower (that is, not in-air) solidifying materials by leveraging alginate as a structural template (fig. S4) (25).

The droplet or particle sizes could be tuned by more than an order of magnitude by varying the nozzle diameter and actuation frequency (Fig. 3G). As shown in Fig. 3 (H to K), monodisperse alginate microgels with diameters ranging from 20 μm (at 0.2 ml/min) to 300 μm (at 6.5 ml/min) were produced by using nozzles with different diameters (overview photos are shown in fig. S3, B and C). The droplet or particle diameter could be fine-tuned by altering the actuation frequency $f$, as shown in Fig. 3K. The probability density function ($P$) of the particle size revealed monodisperse particles with a coefficient of variation of <5% (the SD divided by the mean), as plotted in Fig. 3K and fig. S5.

The particle shape was controlled by altering the velocity ratio between the jets, as shown in Fig. 3 (L to N). Increasing the velocity of the intact jet while maintaining the velocity of the droplets resulted in the formation of elongated particles, as shown in Fig. 3 (M and N) (details are provided in section S1 and fig. S6). Microfibers were produced by operating the same setup in the jet-jet mode (Fig. 3O), as enabled by simply moving the nozzles closer to each other. First, fibers of homogeneous thickness were produced as shown in Fig. 3P, where the cross-linker solution impacts with the gel precursor solution close to nozzle 1. With nozzle actuation turned on while moving the jet's impact location closer to the breakup point (that is, $L \to L_B$), we could produce fibers with periodic thickenings (that is, "beaded fibers"), as shown in Fig. 3Q and fig. S3 (D to F).

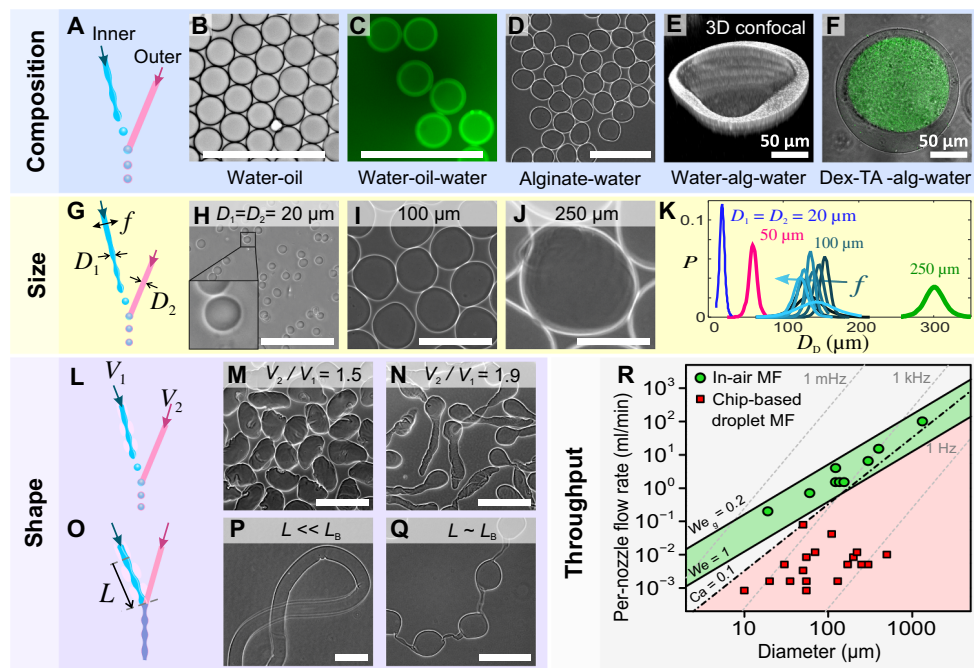

**Fig. 3. IAMF enables high-throughput production of monodisperse microemulsions and microsuspensions with various compositions, sizes, and shapes.** Schematic diagrams (left) indicate the relevant control parameters. (**A** to **F**) IAMF operated in "drop-jet" mode enabled the production of monodisperse (B) w/o emulsions, (C) double emulsions, (D) spherical particle suspensions, and (E) single-material and (F) multimaterial core-shell particles. alg, alginate; dex, dextran; TA, tyramine. (**G** to **K**) Tuning the microparticle size. (H to J) Particles produced using nozzle diameters of 20, 100, and 250 μm, respectively. (K) Probability $P$ of the particle size as a function of nozzle diameter (indicated per curve) and actuation frequency. Colors from black to pale blue indicate increasing actuation frequencies of 2.3, 3.5, 4, 4.5, 5, 6, 7, and 8 kHz. (**L** to **N**) Elongated particles were made by increasing the relative jet velocity. (**O**) IAMF operated in "jet-jet" mode enabled the production of (**P**) straight and (**Q**) pearl-lace morphologies. (**R**) Throughput as a function of nozzle diameter for IAMF and chip-based droplet microfluidics (MF). The maximum per-nozzle throughput of monodispersed droplet production using chip-based microfluidics is limited by Ca = 0.1 and We = 1. The production throughput window of IAMF is determined by We$_{ej}$ = 1 (that is, minimum) and We$_g$ = 0.2 (that is, maximum). Green circles are data points obtained using our IAMF setup. Red squares are data points obtained from previously reported studies on droplet microfluidics (2, 44–52). Droplet production frequencies are indicated with gray dashed lines. Scale bars, 200 μm (unless otherwise indicated).

The diameter and throughput of IAMF-based droplet and particle generation are compared to conventional chip-based droplet microfluidics in Fig. 3R. Because IAMF is based on jetting, the lower flow rate of IAMF is bounded by the ejection Weber number $We_{ej} = \rho_1 V_1^2 D_1 / \sigma_1 > 1$ (fig. S2). The upper production rate of IAMF is presumably limited by wind-induced breakup of the jet or droplet train, which occurs for gas Weber numbers $We_g = \rho_g / \rho We_{ej} > 0.2$, where $\rho_g$ is the density of the gas (26). In contrast, the production of monodisperse droplets using microfluidic chips requires operation within the squeezing or dripping regime, which is bound by $We_{ej} \lesssim 1$ and capillary numbers $Ca = \mu_c V_c / \sigma \lesssim 0.1$, where $\mu_c$ and $V_c$ denote the outer phase's viscosity and velocity, respectively, and $\sigma$ denotes the interfacial tension between the liquids (14). These constraints imply that IAMF is intrinsically faster than chip-based droplet microfluidics. The production rates of IAMF are compared with existing conventional microfluidic droplet generators confirmed in Fig. 3R, revealing that IAMF is typically 100 times faster than chip-based droplet microfluidics. Hence, a single IAMF nozzle is able to produce droplets at similar production rate as compared to an up-scaled microfluidic chip consisting of 364 parallelized droplet generators (27). In section S3, we concisely compare IAMF to alternative droplet and particle production platforms.

### Additive manufacturing of modular (bio)materials

Through direct deposition of in-air formed particles or capsules onto a substrate, IAMF enables printing of 3D multiscale modular materials in one step. To validate this approach, an alginate jet with reduced surface tension was impacted onto $CaCl_2$-containing droplets, resulting in a stream of shape-stable core-shell alginate particles, as shown in Fig. 3 (E and F). Upon deposition onto a substrate, these soft particles stick to the construct without entraining any visible air bubbles and provide sufficient structural support for 3D freeforms (Fig. 4A). As an example, we created a hollow hydrogel cylinder by the directed deposition of such core-shell particles onto a rotating substrate (Fig. 4B and movie S1). The microscale architecture of the modular freeform could be altered by tuning the microparticle composition. For example, single-material core-shell particles formed a liquid-filled foam (Fig. 4C), whereas multimaterial core-shell particles formed a multimaterial solid construct (Fig. 4D). Shape-stable constructs could also be formed onto substrates with an arbitrary inclination angle—and even upside-down—as demonstrated by omnidirectional deposition using a handheld IAMF device (fig. S7 and movie S2).

Furthermore, IAMF could be readily used for one-step generation of injectable modular materials by combining rapidly (that is, in-air) solidifying droplet cores and slowly (that is, after injection) solidifying droplet shells, as shown in Fig. 4E. Upon deposition, particles or fibers were lubricated by their still liquid shell that solidifies after the mold has been filled. As an example, we filled a bone-shaped mold with in-air formed particles (Fig. 4, F to H) and released the intact construct after cross-linking. The micro- and mesoscale architectures consisted of stem cells encapsulated by alginate particles that were embedded in a dextran-based hydrogel matrix, respectively (Fig. 4, G and H). Alternative microarchitectures can be readily produced by, for example, tuning the building block's shape into a fiber, as shown in fig. S8.

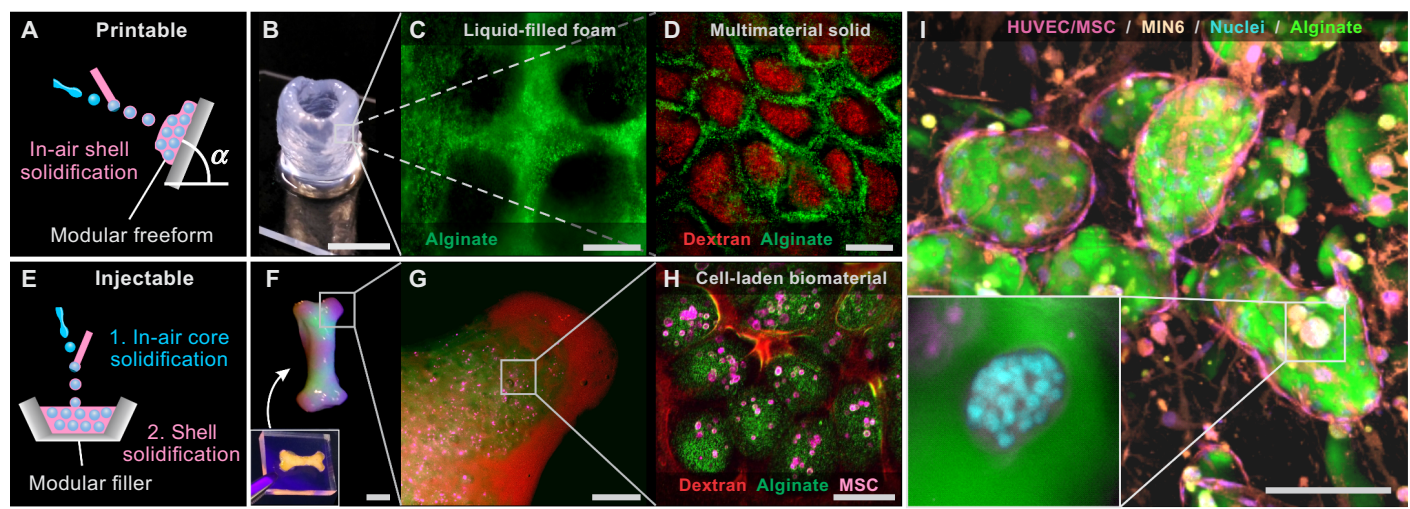

**Fig. 4. One-step additive manufacturing and injection molding of 3D multiscale modular (bio)materials.** (**A**) Modular freeforms with a controlled microarchitecture were manufactured by stacking of shape-stable core-shell particles. (**B** to **D**) A hollow cylinder was formed by deposition of the composite jet onto a rotating substrate. By altering the building blocks' composition, the resulting microarchitecture consisted of (C) a liquid-filled foam or (D) a multimaterial modular solid, where the cross-linker for the core was added to the shell and vice versa. (**E**) To eject a modular filler, only the droplets' cores are solidified in the air, whereas the slower solidifying shells enable seamless filling of the mold. (**F** to **H**) A modular construct was produced by filling a bone-shaped mold. Inset: Hydrogel construct while still in the mold. The 3D multiscale modular material consisted of MSCs (pink), encapsulated in alginate microspheres (green) that are embedded in dextran-tyramine hydrogel (red). (**I**) Injection-molded multiscale modular tissue construct with optimized cellular micro- and macroenvironments. The construct consisted of insulin-producing pancreatic β cells (MIN6; beige with blue nuclei) that were encapsulated in alginate microparticles (green). The cell-laden microparticles were encapsulated within a proangiogenic fibrin network that contained human endothelial and stem cells (pink with blue nuclei). The microenvironments supported MIN6 cell proliferation, whereas the macroenvironment supported the formation of an endothelial cellular network within 7 days of in vitro culture. HUVEC, human umbilical cord endothelial cell. Scale bars, 1 cm (B and F), 5 mm (G), and 100 μm (C, D, H, and I).

Bottom-up module-based additive manufacturing has particular relevance for tissue engineering because it is an effective approach to build constructs that mimic the intricate multiscale microarchitecture of native tissues (28–30). To investigate IAMF's potential in this regard, we analyzed both the effect of cell encapsulation on particle formation (fig. S9) and the influence of IAMF-based processing on the cells (fig. S10). First, the size and shape of alginate microgels were observed to be robust for concentrations of human mesenchymal stem cells (MSCs) up to $10^6$ cells/ml, and the amount of cells per capsule followed the Poisson distribution (fig. S9) (31). For higher cell concentrations, the particles became more polydisperse and larger because the jet's breakup was affected by the incorporated cells. Second, the influence of IAMF on the cells was assessed in fig. S10. More than 90% of the MSCs that had been encapsulated using 100-μm nozzles at a rate of 1.2 ml/min remained viable during at least 1 week of in vitro culture, irrespective of the nozzle-collector distance (fig. S10B). When increasing the per-nozzle flow rate to 2 ml/min, the 1-week cell survival remained over 80%. Moreover, the encapsulated MSCs remained functional, as indicated by their maintained adipogenic differentiation capacity (fig. S10C). Building on these promising results, we engineered a multicellular and multimaterial 3D tissue construct, as shown in Fig. 4L. Insulin-producing pancreatic β cells (that is, MIN6) were encapsulated in alginate microparticles that were surrounded by a coculture of human endothelial cells and MSCs embedded in a proangiogenic fibrin gel. Within 1 week, proliferating insulin-positive MIN6 cells formed cell aggregates within the alginate microenvironments, whereas endothelial and stem cells organized into a prevascular network that stained positive for von Willebrand factor and permeated throughout the fibrin gel (fig. S10). Hence, IAMF enables rapid one-step manufacturing of 3D modular cell-laden biomaterials with distinct cellular micro- and macroenvironments of a clinically relevant size, which is a promising yet challenging direction in the field of tissue engineering (29, 32–34).

## DISCUSSION

Here, we presented IAMF, a platform for the rapid production of droplets, particles, and fibers with controlled composition, shape, and size, as demonstrated in Fig. 3 (a schematic overview is provided in fig. S11A). The per-nozzle flow rates of IAMF exceed conventional chip-based microfluidics by typically two orders of magnitude. IAMF also omits the need for cleanroom-based chip fabrication and channel wall surface treatments, prevents solidification-induced clogging, and allows oil-free manufacturing of microparticles (for example, microgels) (11–13, 15). These characteristics facilitate the application of microfluidic technologies in environments that are not readily compatible with microfluidic chips (11). For example, rapid oil-free production of cell-laden hydrogel particles would benefit cell microencapsulation strategies for in situ use and therefore their clinical translation. Furthermore, IAMF eliminates the need for high voltages or rotating equipment to maintain droplet separation, which are mandatory in existing jet-based methods to produce complex monodisperse suspensions and emulsions (see section S3). This feat is achieved by replacing cross-linking in a collector bath by in-line solidification. Eventually, integration of IAMF with existing jet-based technologies such as fluorescence-activated cell sorters (35) and liquid sheet–based approaches (36–38) may expand the diversity of particle compositions even further.

IAMF also enables additive manufacturing of multiscale and functional materials in one step by controlled in-air solidification of microdroplets and their subsequent deposition onto a substrate. This ability contrasts chip-based microfluidics, which requires a nonsolidifying coflow, as well as current additive manufacturing methods as described in section S4. A wider variety of construct shapes could be achieved by, for example, mounting IAMF nozzles onto automated xyz stages, whereas intricate details could be provided by placement of individual droplets by integration with, for example, inkjet printing (26, 39, 40). Finally, IAMF combines shape fidelity, high throughput, and cytocompatibility in the additive manufacturing of cell-containing materials. This combination has remained elusive in biofabrication technologies (41) but is now enabled by decoupling the material properties at the nozzle from those at the substrate. Low-viscosity inks can be used to form shape-stable particles on-the-fly, which can be stacked into larger solid constructs. This low viscosity reduces shear stresses at the nozzle and therefore prevents stress-induced cell death, as discussed in section S4. Finally, IAMF can be adapted to enable handheld device operation, as pioneered in this work. Especially tissue engineers and surgeons may benefit from this ability because current methods for in situ repair of wounds and defects (42) are expected to benefit from monodisperse modular hydrogel filling strategies.

## MATERIALS AND METHODS
### Device preparation and operation

Liquid jets were ejected from fused silica tubing (IDEX Health & Science) with an outer diameter of 360 μm and inner diameters of 20 ± 1, 50 ± 3, 100 ± 3, 150 ± 5, or 250 ± 5 μm. Because long sections of tubing with small diameters result in a high pressure drop (described by the Darcy-Weisbach equation: $\Delta P = \frac{128}{\pi} \cdot \frac{\mu Q L_t}{D_t^4}$, where $Q$ is the flow rate, $D_t$ is the tip inner diameter, and $L_t$ is the length of the tip), stalling of the syringe pumps was initially observed for the nozzles with diameters of <100 μm. Therefore, nozzles were cut using a Shortix capillary cutter (SGT) and glued into PEEK tubing (IDEX Health & Science) with a larger inner diameter of 0.5 mm and an outer diameter of 1.59 mm ($^1/_{16}$ inch) using a quick-set epoxy adhesive (RS 850-956, RS Components Ltd.). The length of these nozzles is 4 ± 1 mm, corresponding to the smallest length that could be manually cut and glued. Their semitransparent end and the transition to the PEEK tubing (in black) are just visible in Fig. 2A. The PEEK tubing was mounted onto a piezoelectric actuator using two-sided tape (3M) and standard optical components (Thorlabs). The piezo was actuated using a 150-V sine wave, and its vibration results in translation of the nozzle tip perpendicular to the flow direction.

When starting up, a few dripping droplets formed at each nozzle before the establishment of steady jets or droplet train. Dripping of alginate from nozzle 1 onto the $CaCl_2$-containing nozzle 2 was observed to result in clogging of nozzle 2, especially for the smaller (20 μm) nozzles because these were placed close (<1 mm) to each other when operating the setup. The diameter of these dripping droplets was approximately 3 mm. Therefore, a horizontal separation of >3 mm was applied before each experiment; the nozzles were subsequently aligned when stable jets had formed.

The dynamics of jet breakup and impact were continuously monitored with a previously described stroboscopic visualization setup (39). In short, an Nd:YAG laser with a pulse duration of 6 ns was directed onto a fluorescent diffuser to generate a homogenous background illumination. A camera (PCO Sensicam QE) with a shutter time of 400 ns was synchronized with these flashes to capture high-resolution snapshots of the experiments. The camera was mounted on a separate xyz stage to facilitate alignment with the in-air impact location.

Both nozzles (as required for the two jets or droplet trains) were of equal diameter and operated at equal flow rates except for the elongated particle experiments. The velocities of the jets were calculated by volume conservation: $Q = \pi \left(\frac{D}{2}\right)^2 V$. The respective position of the nozzles was controlled by mounting one of the nozzles onto a 3D stage with 1-μm precision (Thorlabs). In the handheld device, a screw was used to deflect the nozzle tip. For both devices, the jets were aligned by moving one of them in the z direction by turning a screw until controlled breakup of both jets was observed with the camera (representative images are shown in fig. S2, G to I). After this calibration, the flow generally remained stable for the duration of the experiment (~3 min), as monitored with the camera. Over this time scale, $10^5$ to $10^7$ droplets were produced, depending on the nozzles' diameter.

To control the flow rate, a standard syringe pump (type PhD 2000, Harvard Apparatus) and plastic syringes were used (5 or 10 ml; Luer-Lok, BD). A high-power syringe pump (Harvard Apparatus) and steel syringes (9 ml; Luer-Lok, Harvard Apparatus) were used in case excessive pressure drops over the nozzle tip caused the standard syringe pump to stall (that is, mainly for the 20-μm nozzles). Threaded adapters (IDEX Health & Science) were used to connect the syringes to the PEEK tubing in which the nozzle tips were glued as described.

The control parameters are detailed in table S1. To create various materials, three liquids were generally used: Liquid 1 constitutes the jet or droplet train (the core), liquid 2 constitutes the jet (shell), and liquid 3 constitutes the bath (not applicable to printing as shown in Fig. 4 and fig. S8). All aqueous solutions were prepared using water, phosphate-buffered saline (PBS), or cell culture medium. The setup was operated at ejection Weber number $10 < We_{ej} < 32$, for which controlled ejection and breakup were observed. The corresponding impact Weber numbers were typically 3.5 and all below 10. For visualization purposes, <0.1% of dextran–fluorescein isothiocyanate (2000 kDa; Sigma-Aldrich), rhodamine B dye, or rhodamine B–stained particles (diameter, 500 nm) were added to the ejected liquids.

The surface tension of ethanol-containing $CaCl_2$ solutions was measured using the hanging drop method on an optical contact angle measuring system (OCA 15Pro, DataPhysics). The results overlap with previously reported measurements of ethanol/water mixtures within the experimental error (5%), indicating that $CaCl_2$ had no significant effect on the surface tension (fig. S12).

### Cell Isolation, expansion, and encapsulation

Human MSCs were isolated from fresh bone marrow samples and cultured as previously described. The use of patient material was approved by the local ethical committee of the Medisch Spectrum Twente (Enschede, Netherlands), and informed written consent was obtained for all samples. In short, nucleated cells in the bone marrow aspirates were counted, seeded in tissue culture flasks at a density of $5 \times 10^5$ cells/cm$^2$, and cultured in MSC proliferation medium, consisting of 10% (v/v) fetal bovine serum (FBS; Sigma-Aldrich), penicillin (100 U/ml) with streptomycin (100 μg/ml) (Gibco), 1% (v/v) GlutaMAX (Gibco), 0.2 mM ascorbic acid (Sigma-Aldrich), and basic fibroblast growth factor (1 ng/ml) (ISOKine bFGF, Neuromics; added fresh) in α-minimum essential medium with nucleosides (Gibco). Cells were cultured under 5% $CO_2$ at 37°C, and the medium was replaced two to three times per week. When cell culture reached near-confluence, the cells were detached using 0.25% (w/v) trypsin-EDTA (Gibco) at 37°C and subsequently subcultured or used for experimentation. For cell encapsulation, MSCs were suspended in MSC proliferation medium and mixed with 1% (w/v) sodium alginate (80 to 120 cP; Wako Chemicals) in PBS (Gibco) in a 1:1 ratio. The cell-laden hydrogel precursor solution was loaded into a disposable syringe and connected to the IAMF setup for encapsulation. After encapsulation, cell-laden microgels were cultured in six-well plates (Nunc) with MSC proliferation medium under 5% $CO_2$ at 37°C, which was refreshed three times per week. The viability of encapsulated MSCs was analyzed using a live/dead assay (Molecular Probes) following the manufacturer's protocol and visualization using a fluorescence microscope (EVOS FL, Thermo Fisher Scientific). Images were analyzed using ImageJ software, and cell viability was quantified via manual counting. Endothelial cell-laden modular constructs were formed by adding a third jet to the system, which contained fibrin precursor solution and thrombin solution (50 U/ml) (Sigma-Aldrich) that were mixed immediately before jetting using a T-junction. Fibrin precursor solution was prepared by suspending HUVECs and MSCs into endothelial growth medium-2 (EGM-2) without FBS supplemented with fibrinogen (10 mg/ml) (Sigma-Aldrich). Just before producing the constructs, 5% (v/v) FBS was added to the fibrin precursor solution as previously described (43). After 5-min incubation at room temperature, the constructs were incubated for 20 min at 37°C to complete polymerization, after which a 1:1 mixture of MIN6 proliferation medium and EGM-2 was added on top. The constructs were cultured for 1 week, and the medium was refreshed every 2 to 3 days. The constructs were then fixated using 4% (w/v) formaldehyde (Sigma-Aldrich), permeabilized using 0.1% (v/v) Triton X-100 (Sigma-Aldrich), blocked using 10% (w/v) bovine serum albumin (Sigma-Aldrich), and stained using 1:100 anti-CD31 (AB32457, Abcam) and 1:100 anti-insulin (AB7842, Abcam), in combination with 1:400 Alexa Fluor 488–, tetramethyl rhodamine isothiocyanate (TRITC)–, or Alexa Fluor 647–labeled secondary antibodies, and 4′,6-diamidino-2-phenylindole (DAPI) as counter staining. All staining solutions were prepared using Hanks' balanced salt solution (Sigma-Aldrich), as PBS dissolves the alginate. Alternatively, constructs were impregnated in Cryomatrix (Shandon), cryosectioned (7 μm, Leica cryostat), and stained as described. Subsequent imaging was performed using a fluorescence confocal microscope (Nikon A1+).

### SUPPLEMENTARY MATERIALS

Supplementary material for this article is available at http://advances.sciencemag.org/cgi/content/full/4/1/eaao1175/DC1
section S1. Rayleigh-Plateau breakup
section S2. Shape control
section S3. Comparing IAMF to droplet-based particle formation technologies
section S4. IAMF with respect to additive manufacturing technologies
fig. S1. Schematic of the setup.
fig. S2. Ejection and breakup regimes.
fig. S3. Overview of particles and fibers.
fig. S4. Hydrogel templating.
fig. S5. Particle size distributions.
fig. S6. Elongation of droplets and particles.
fig. S7. IAMF handheld device.
fig. S8. Fiber-based modular materials.
fig. S9. Characterization of IAMF-based cell microencapsulation.
fig. S10. Viability and function of IAMF-processed cells.
fig. S11. Overview of IAMF-based materials and throughput.
fig. S12. Surface tension of ethanol/water mixtures.

movie S1. One-step 3D modular printing a solid freeform.
movie S2. Omnidirectional printing using IAMF handheld device.
table S1. Detailed process parameters of key experiments.
References (53–83)

## REFERENCES AND NOTES

1. R. Seemann, M. Brinkmann, T. Pfohl, S. Herminghaus, Droplet based microfluidics. *Rep. Prog. Phys.* **75**, 016601 (2012).
2. D. Dendukuri, P. S. Doyle, The synthesis and assembly of polymeric microparticles using microfluidics. *Adv. Mater.* **21**, 4071–4086 (2009).
3. H. Onoe, T. Okitsu, A. Itou, M. Kato-Negishi, R. Gojo, D. Kiriya, K. Sato, S. Miura, S. Iwanaga, K. Kuribayashi-Shigetomi, Y. T. Matsunaga, Y. Shimoyama, S. Takeuchi, Metre-long cell-laden microfibres exhibit tissue morphologies and functions. *Nat. Mater.* **12**, 584–590 (2013).
4. J.-W. Kim, A. S. Utada, A. Fernández-Nieves, Z. Hu, D. A. Weitz, Fabrication of monodisperse gel shells and functional microgels in microfluidic devices. *Angew. Chem. Int. Ed. Engl.* **46**, 1819–1822 (2007).
5. G. M. Whitesides, The origins and the future of microfluidics. *Nature* **442**, 368–373 (2006).
6. T. A. Duncombe, A. M. Tentori, A. E. Herr, Microfluidics: Reframing biological enquiry. *Nat. Rev. Mol. Cell Biol.* **16**, 554–567 (2015).
7. S. Ma, J. Thiele, X. Liu, Y. Bai, C. Abell, W. T. S. Huck, Fabrication of microgel particles with complex shape via selective polymerization of aqueous two-phase systems. *Small* **8**, 2356–2360 (2012).
8. E. Kang, G. S. Jeong, Y. Y. Choi, K. H. Lee, A. Khademhosseini, S.-H. Lee, Digitally tunable physicochemical coding of material composition and topography in continuous microfibres. *Nat. Mater.* **10**, 877–883 (2011).
9. G. Y. Huang, L. H. Zhou, Q. C. Zhang, Y. M. Chen, W. Sun, F. Xu, T. J. Lu, Microfluidic hydrogels for tissue engineering. *Biofabrication* **3**, 012001 (2011).
10. S. Xu, Z. Nie, M. Seo, P. Lewis, E. Kumacheva, H. A. Stone, P. Garstecki, D. B. Weibel, I. Gitlin, G. M. Whitesides, Generation of monodisperse particles by using microfluidics: Control over size, shape, and composition. *Angew. Chem. Int. Ed. Engl.* **44**, 724–728 (2005).
11. L. R. Volpatti, A. K. Yetisen, Commercialization of microfluidic devices. *Trends Biotechnol.* **32**, 347–350 (2014).
12. S. Mashaghi, A. Abbaspourrad, D. A. Weitz, A. M. van Oijen, Droplet microfluidics: A tool for biology, chemistry and nanotechnology. *Trends Anal. Chem.* **82**, 118–125 (2016).
13. J. H. Kim, T. Y. Jeon, T. M. Choi, T. S. Shim, S.-H. Kim, S.-M. Yang, Droplet microfluidics for producing functional microparticles. *Langmuir* **30**, 1473–1488 (2014).
14. J. K. Nunes, S. S. H. Tsai, J. Wan, H. A. Stone, Dripping and jetting in microfluidic multiphase flows applied to particle and fibre synthesis. *J. Phys. D Appl. Phys.* **46**, 114002 (2013).
15. S. H. Ching, N. Bansal, B. Bhandari, Alginate gel particles—A review of production techniques and physical properties. *Crit. Rev. Food Sci. Nutr.* **57**, 1133–1152 (2017).
16. R.-H. Chen, S.-L. Chiu, T.-H. Lin, Collisions of a string of water drops on a water jet of equal diameter. *Exp. Therm. Fluid Sci.* **31**, 75–81 (2006).
17. S. Wildeman, C. W. Visser, C. Sun, D. Lohse, On the spreading of impacting drops. *J. Fluid Mech.* **805**, 636–655 (2016).
18. D. G. A. L. Aarts, H. N. W. Lekkerkerker, H. Guo, G. H. Wegdam, D. Bonn, Hydrodynamics of droplet coalescence. *Phys. Rev. Lett.* **95**, 164503 (2005).
19. F. Blanchette, Simulation of mixing within drops due to surface tension variations. *Phys. Rev. Lett.* **105**, 074501 (2010).
20. R.-H. Chen, Diesel–diesel and diesel–ethanol drop collisions. *Appl. Therm. Eng.* **27**, 604–610 (2007).
21. C. Planchette, E. Lorenceau, G. Brenn, Liquid encapsulation by binary collisions of immiscible liquid drops. *Colloids Surf. A Physicochem. Eng. Aspects* **365**, 89–94 (2010).
22. B. F. van Capelleveen, R. B. J. Koldeweij, D. Lohse, C. W. Visser, On the universality of Marangoni-driven spreading along liquid-liquid interfaces. http://arxiv.org/abs/1712.03192 (2017).
23. G. Skjåk-Bræk, H. Grasdalen, O. Smidsrød, Inhomogeneous polysaccharide ionic gels. *Carbohydr. Polym.* **10**, 31–54 (1989).
24. J. Zhang, C. R. Daubert, E. A. Foegeding, Fracture analysis of alginate gels. *J. Food Sci.* **70**, e425–e431 (2005).
25. A. Tamayol, A. H. Najafabadi, B. Aliakbarian, E. Arab-Tehrany, M. Akbari, N. Annabi, D. Juncker, A. Khademhosseini, Hydrogel templates for rapid manufacturing of bioactive fibers and 3D constructs. *Adv. Healthc. Mater.* **4**, 2146–2153 (2015).
26. W. van Hoeve, S. Gekle, J. H. Snoeijer, M. Versluis, M. P. Brenner, D. Lohse, Breakup of diminutive Rayleigh jets. *Phys. Fluids* **22**, 122003 (2010).
27. A. Ofner, D. G. Moore, P. A. Rühs, P. Schwendimann, M. Eggersdorfer, E. Amstad, D. A. Weitz, A. R. Studart, High-throughput step emulsification for the production of functional materials using a glass microfluidic device. *Macromol. Chem. Phys.* **218**, 1600472 (2017).
28. J. S. Liu, Z. J. Gartner, Directing the assembly of spatially organized multicomponent tissues from the bottom up. *Trends Cell Biol.* **22**, 683–691 (2012).
29. J. Leijten, A. Khademhosseini, From nano to macro: Multiscale materials for improved stem cell culturing and analysis. *Cell Stem Cell* **18**, 20–24 (2016).
30. J. W. Nichol, A. Khademhosseini, Modular tissue engineering: Engineering biological tissues from the bottom up. *Soft Matter* **5**, 1312–1319 (2009).
31. D. J. Collins, A. Neild, A. DeMello, A.-Q. Liu, Y. Ai, The Poisson distribution and beyond: Methods for microfluidic droplet production and single cell encapsulation. *Lab Chip* **15**, 3439–3459 (2015).
32. S. M. Oliveira, R. L. Reis, J. F. Mano, Towards the design of 3D multiscale instructive tissue engineering constructs: Current approaches and trends. *Biotechnol. Adv.* **33**, 842–855 (2015).
33. T. Kamperman, S. Henke, A. van den Berg, S. R. Shin, A. Tamayol, A. Khademhosseini, M. Karperien, J. Leijten, Single cell microgel based modular bioinks for uncoupled cellular micro- and macroenvironments. *Adv. Healthc. Mater.* **6**, 1600913 (2016).
34. T. Kamperman, S. Henke, C. W. Visser, M. Karperien, J. Leijten, Centering single cells in microgels via delayed crosslinking supports long-term 3D culture by preventing cell escape. *Small* **13**, 1603711 (2017).
35. L. A. Herzenberg, R. G. Sweet, Fluorescence-activated cell sorting. *Sci. Am.* **234**, 108–117 (1976).
36. R. Houben, thesis, University of Twente (2012).
37. N. Blanco-Pascual, R. B. J. Koldeweij, R. S. A. Stevens, M. P. Montero, M. C. Gómez-Guillén, A. T. Ten Cate, Peptide microencapsulation by core–shell printing technology for edible film application. *Food Bioprocess Technol.* **7**, 2472–2483 (2014).
38. A. Walther, A. H. E. Müller, Janus particles: Synthesis, self-assembly, physical properties, and applications. *Chem. Rev.* **113**, 5194–5261 (2013).
39. C. W. Visser, P. E. Frommhold, S. Wildeman, R. Mettin, D. Lohse, C. Sun, Dynamics of high-speed micro-drop impact: Numerical simulations and experiments at frame-to-frame times below 100 ns. *Soft Matter* **11**, 1708–1722 (2015).
40. H. Wijshoff, The dynamics of the piezo inkjet printhead operation. *Phys. Rep.* **491**, 77–177 (2010).
41. J. Malda, J. Visser, F. P. Melchels, T. Jüngst, W. E. Hennink, W. J. A. Dhert, J. Groll, D. W. Hutmacher, 25th anniversary article: Engineering hydrogels for biofabrication. *Adv. Mater.* **25**, 5011–5028 (2013).
42. J. Hendriks, C. W. Visser, S. Henke, J. Leijten, D. B. F. Saris, C. Sun, D. Lohse, M. Karperien, Optimizing cell viability in droplet-based cell deposition. *Sci. Rep.* **5**, 11304 (2015).
43. S. K. Both, A. J. C. van der Muijsenberg, C. A. van Blitterswijk, J. de Boer, J. D. de Bruijn, A rapid and efficient method for expansion of human mesenchymal stem cells. *Tissue Eng.* **13**, 3–9 (2007).
44. T. Femmer, A. Jans, R. Eswein, N. Anwar, M. Moeller, M. Wessling, A. J. C. Kuehne, High-throughput generation of emulsions and microgels in parallelized microfluidic drop-makers prepared by rapid prototyping. *ACS Appl. Mater. Interfaces* **7**, 12635–12638 (2015).
45. E. Tumarkin, L. Tzadu, E. Csaszar, M. Seo, H. Zhang, A. Lee, R. Peerani, K. Purpura, P. W. Zandstra, E. Kumacheva, High-throughput combinatorial cell co-culture using microfluidics. *Integr. Biol.* **3**, 653–662 (2011).
46. E. W. M. Kemna, R. M. Schoeman, F. Wolbers, I. Vermes, D. A. Weitz, A. van den Berg, High-yield cell ordering and deterministic cell-in-droplet encapsulation using Dean flow in a curved microchannel. *Lab Chip* **12**, 2881–2887 (2012).
47. L. Yobas, S. Martens, W.-L. Ong, N. Ranganathan, High-performance flow-focusing geometry for spontaneous generation of monodispersed droplets. *Lab Chip* **6**, 1073–1079 (2006).
48. K. Liu, H.-J. Ding, J. Liu, Y. Chen, X.-Z. Zhao, Shape-controlled production of biodegradable calcium alginate gel microparticles using a novel microfluidic device. *Langmuir* **22**, 9453–9457 (2006).
49. H. Zhang, E. Tumarkin, R. Peerani, Z. Nie, R. M. A. Sullan, G. C. Walker, E. Kumacheva, Microfluidic production of biopolymer microcapsules with controlled morphology. *J. Am. Chem. Soc.* **128**, 12205–12210 (2006).
50. Y.-S. Lin, C.-H. Yang, Y.-Y. Hsu, C.-L. Hsieh, Microfluidic synthesis of tail-shaped alginate microparticles using slow sedimentation. *Electrophoresis* **34**, 425–431 (2013).
51. S. Utech, R. Prodanovic, A. S. Mao, R. Ostafe, D. J. Mooney, D. A. Weitz, Microfluidic generation of monodisperse, structurally homogeneous alginate microgels for cell encapsulation and 3D cell culture. *Adv. Healthc. Mater.* **4**, 1628–1633 (2015).
52. A. S. Utada, A. Fernandez-Nieves, H. A. Stone, D. A. Weitz, Dripping to jetting transitions in coflowing liquid streams. *Phys. Rev. Lett.* **99**, 094502 (2007).
53. J. Eggers, E. Villermaux, Physics of liquid jets. *Rep. Prog. Phys.* **71**, 036601 (2008).
54. H. Brandenberger, D. Nüssli, V. Piëch, F. Widmer, Monodisperse particle production: A method to prevent drop coalescence using electrostatic forces. *J. Electrostat.* **45**, 227–238 (1999).
55. A. Martinsen, G. Skjåk-Braek, O. Smidsrød, Alginate as immobilization material: I. Correlation between chemical and physical properties of alginate gel beads. *Biotechnol. Bioeng.* **33**, 79–89 (1989).

56. A. Choi, K. D. Seo, D. W. Kim, B. C. Kim, D. S. Kim, Recent advances in engineering microparticles and their nascent utilization in biomedical delivery and diagnostic applications. *Lab Chip* **17**, 591–613 (2017).
57. J. Xie, J. Jiang, P. Davoodi, M. P. Srinivasan, C.-H. Wang, Electrohydrodynamic atomization: A two-decade effort to produce and process micro-/nanoparticulate materials. *Chem. Eng. Sci.* **125**, 32–57 (2015).
58. V.-T. Tran, J.-P. Benoît, M.-C. Venier-Julienne, Why and how to prepare biodegradable, monodispersed, polymeric microparticles in the field of pharmacy? *Int. J. Pharm.* **407**, 1–11 (2011).
59. U. Prüsse, L. Bilancetti, M. Bučko, B. Bugarski, J. Bukowski, P. Gemeiner, D. Lewińska, V. Manojlovic, B. Massart, C. Nastruzzi, V. Nedovic, D. Poncelet, S. Siebenhaar, L. Tobler, A. Tosi, A. Vikartovská, K.-D. Vorlop, Comparison of different technologies for alginate beads production. *Chem. Pap.* **62**, 364–374 (2008).
60. M. Whelehan, I. W. Marison, Microencapsulation by dripping and jet breakup. *Bioencapsulation Innov.* **1**, 4–10 (2011).
61. M. Hayakawa, H. Onoe, K. H. Nagai, M. Takinoue, Complex-shaped three-dimensional multi-compartmental microparticles generated by diffusional and Marangoni microflows in centrifugally discharged droplets. *Sci. Rep.* **6**, 20793 (2016).
62. C. L. Herran, Y. Huang, Alginate microsphere fabrication using bipolar wave-based drop-on-demand jetting. *J. Manuf. Process.* **14**, 98–106 (2012).
63. B. Derby, Inkjet printing of functional and structural materials: Fluid property requirements, feature stability, and resolution. *Annu. Rev. Mater. Res.* **40**, 395–414 (2010).
64. A. Piqué, R. C. Y. Auyeung, H. Kim, N. A. Charipar, S. A. Mathews, Laser 3D micro-manufacturing. *J. Phys. D Appl. Phys.* **49**, 223001 (2016).
65. P. Delaporte, A.-P. Alloncle, Laser-induced forward transfer: A high resolution additive manufacturing technology. *Opt. Laser Technol.* **78**, 33–41 (2016).
66. R. Levato, M. A. Mateos-Timoneda, J. A. Planell, Preparation of biodegradable polylactide microparticles via a biocompatible procedure. *Macromol. Biosci.* **12**, 557–566 (2012).
67. I. Gibson, D. W. Rosen, B. Stucker, *Additive Manufacturing Technologies* (Springer US, 2010), vol. 76.
68. M. Vaezi, H. Seitz, S. Yang, A review on 3D micro-additive manufacturing technologies. *Int. J. Adv. Manuf. Technol.* **67**, 1721–1754 (2013).
69. T. Jungst, W. Smolan, K. Schacht, T. Scheibel, J. Groll, Strategies and molecular design criteria for 3D printable hydrogels. *Chem. Rev.* **116**, 1496–1539 (2016).
70. A. Blaeser, D. F. D. Campos, U. Puster, W. Richtering, M. M. Stevens, H. Fischer, Controlling shear stress in 3D bioprinting is a key factor to balance printing resolution and stem cell integrity. *Adv. Healthc. Mater.* **5**, 326–333 (2015).
71. M. S. Mannoor, Z. Jiang, T. James, Y. L. Kong, K. A. Malatesta, W. O. Soboyejo, N. Verma, D. H. Gracias, M. C. McAlpine, 3D printed bionic ears. *Nano Lett.* **13**, 2634–2639 (2013).
72. A. Russo, B. Y. Ahn, J. J. Adams, E. B. Duoss, J. T. Bernhard, J. A. Lewis, Pen-on-paper flexible electronics. *Adv. Mater.* **23**, 3426–3430 (2011).
73. R. L. Truby, J. A. Lewis, Printing soft matter in three dimensions. *Nature* **540**, 371–378 (2016).
74. T. J. Hinton, Q. Jallerat, R. N. Palchesko, J. H. Park, M. S. Grodzicki, H.-J. Shue, M. R. Ramadan, A. R. Hudson, A. W. Feinberg, Three-dimensional printing of complex biological structures by freeform reversible embedding of suspended hydrogels. *Sci. Adv.* **1**, e1500758 (2015).
75. D. B. Kolesky, K. A. Homan, M. A. Skylar-Scott, J. A. Lewis, Three-dimensional bioprinting of thick vascularized tissues. *Proc. Natl. Acad. Sci. U.S.A.* **113**, 3179–3184 (2016).
76. A. Faulkner-Jones, S. Greenhough, J. A. King, J. Gardner, A. Courtney, W. Shu, Development of a valve-based cell printer for the formation of human embryonic stem cell spheroid aggregates. *Biofabrication* **5**, 015013 (2013).
77. J.-U. Park, M. Hardy, S. J. Kang, K. Barton, K. Adair, D. k. Mukhopadhyay, C. Y. Lee, M. S. Strano, A. G. Alleyne, J. G. Georgiadis, P. M. Ferreira, J. A. Rogers, High-resolution electrohydrodynamic jet printing. *Nat. Mater.* **6** (2007).
78. J. R. H. Shaw-Stewart, T. Mattle, T. K. Lippert, M. Nagel, F. A. Nüesch, A. Wokaun, The fabrication of small molecule organic light-emitting diode pixels by laser-induced forward transfer. *J. Appl. Phys.* **113**, 043104 (2013).
79. B. Zhang, J. He, X. Li, F. Xu, D. Li, Micro/nanoscale electrohydrodynamic printing: From 2D to 3D. *Nanoscale* **8**, 15376–15388 (2016).
80. H. Gudapati, M. Dey, I. Ozbolat, A comprehensive review on droplet-based bioprinting: Past, present and future. *Biomaterials* **102**, 20–42 (2016).
81. C. Clanet, J. C. Lasheras, Transition from dripping to jetting. *J. Fluid Mech.* **383**, 307–326 (1999).
82. P. E. Frommhold, A. Lippert, F. L. Holsteyns, R. Mettin, High-speed monodisperse droplet generation by ultrasonically controlled micro-jet breakup. *Exp. Fluids* **55**, 1716 (2014).
83. G. Vazquez, E. Alvarez, J. M. Navaza, Surface tension of alcohol + water from 20 to 50°C. *J. Chem. Eng. Data* **40**, 611–614 (1995).

**Acknowledgments:** We thank J. Leijten, S. Karpitschka, S. Wildeman, and J. Hendriks for discussions; Y. Zhang, P. Dijkstra, V. de Jong, and R. Wang for contributions to trial experiments; P. Halban (University of Geneva, Switzerland) for providing mouse MIN6-B1 cells; and A. J. S. Renard (Ziekenhuisgroep Twente, Netherlands) for providing bone marrow samples. **Funding:** C.W.V., M.K., and D.L. acknowledge MIRA. T.K., L.P.K., and M.K. also acknowledge the Dutch Arthritis Foundation (grants 12-2-411 and LLP-25). D.L. acknowledge the European Research Council (ERC DDD: 740479; ERC PhysBoil: 267166) and the Dutch Organization for Scientific Research (Spinoza: SPI 69-11). **Author contributions:** Conception and first draft by C.W.V. and T.K. Experiments by C.W.V., T.K., and L.P.K. Data interpretation by all authors. Supervision and revisions by D.L. and M.K. **Competing interests:** C.W.V., T.K., D.L., and M.K. are inventors on a patent based on this work filed by the University of Twente (no. PCT/EP2017/057392; priority date: 30 March 2016). C.W.V. and T.K. have the intention to use this patent to found a startup company. The authors declare no other competing interests. **Data and materials availability:** All data needed to evaluate the conclusions in the paper are present in the paper and/or the Supplementary Materials. Additional data related to this paper may be requested from the authors.

Submitted 15 June 2017
Accepted 3 January 2018
Published 31 January 2018
10.1126/sciadv.aao1175

**Citation:** C. W. Visser, T. Kamperman, L. P. Karbaat, D. Lohse, M. Karperien, In-air microfluidics enables rapid fabrication of emulsions, suspensions, and 3D modular (bio)materials. *Sci. Adv.* **4**, eaao1175 (2018).